\documentclass{pnastwo}

\usepackage{graphicx}
\usepackage{amssymb,amsfonts,amsmath}
\usepackage[normalem]{ulem}

\url{www.pnas.org/cgi/doi/10.1073/pnas.0709640104}
\copyrightyear{2008}
\issuedate{Issue Date}
\volume{Volume}
\issuenumber{Issue Number}
\begin{document}


\title{Fluctuations and Shape of Cooperative Rearranging Regions  in Glass-Forming Liquids}

\author{
Giulio Biroli\affil{1}{Institut Physique Th\'eorique (IPhT) CEA Saclay, and CNRS URA 2306, 91191 Gif Sur Yvette,}
Chiara Cammarota\affil{2}{Dipartimento di Fisica, Ed. Marconi, "Sapienza" Universit\`a di Roma, P.le A. Moro 2, 00185 Roma Italy}}

\contributor{Submitted to Proceedings of the National Academy of Sciences
of the United States of America}

\significancetext{Cooperative rearranging regions (CRR) represent a fundamental ingredient of the theory of glass-forming liquids. In recent years, theoretical, numerical and experimental studies determined their spatial extent and its variation approaching the glass transition. In this work we characterize their shape.  
We show that when temperature is lowered, concomitantly to the growth of the size of CRRs, the boundary of CRRs becomes rougher and more fluctuating. We reveal that two distinct static lengths accompany the glass transition: 
the point-to-set and the wandering length. They respectively measure the spatial extent and the fluctuations of the shape of CRRs.
}
\maketitle

\begin{article}
\begin{abstract}
We develop a theory of amorphous interfaces in glass-forming liquids. We show that the statistical properties of these surfaces, which separate regions characterized by different amorphous arrangements of particles, coincide with the ones of domain walls in the random field Ising model.
A major consequence of our results is that super-cooled liquids are characterized by two different static lengths: the point-to-set $\xi_{PS}$ which is a measure of the spatial extent of cooperative rearranging regions and the wandering length $\xi_\perp$ which is related to the fluctuations of their shape. We find that $\xi_\perp$ grows when approaching the glass transition but slower than $\xi_{PS}$. The wandering length increases as $s_c^{-1/2}$, where $s_c$ is the configurational entropy. Our results 
strengthen the relationship with the random field Ising model found in recent works. They are in agreement with previous numerical studies of amorphous interfaces and provide a theoretical framework for explaining numerical and experimental findings on pinned particle systems and static lengths in glass-forming liquids. 

\end{abstract}

\section{Introduction}
Developing a theory of the glass transition remains one of the most fundamental challenge 
of statistical physics and condensed matter. The interest in this problem actually goes well beyond
the physics of molecular super-cooled liquids. The reason is that glassy behaviour is ubiquitous; it appears 
in a large variety of contexts: from physical systems like colloids and granular material to central problems 
in other branches of sciences like computer science, economics and biology. 
Recent years have witnessed important and substantial progress in its understanding. 
Several theoretical approaches have grown in importance and in the level of detailed predictions and 
explanations \cite{birgar13}. In particular, the Random First Order Transition (RFOT) theory originally introduced by Kirkpatrick, Thirumalai and Wolynes \cite{kithwo89}  has been boosted from new theoretical ideas and techniques \cite{mezpar99,boubir04,wollub12} and innovative simulation studies \cite{cagrve07,bibcgv08,parisi09,sautar10,chchta12,berkob12,homare12}. 
The initial idea of Kirkpatrick, Thirumalai and Wolynes that super-cooled liquids are in a 
mosaic state, a kind of micro phase separated state in which the number of possible phases is huge, has been made concrete and testable in analytical computations and numerical simulations.  
We understand now how to define and---measure---the spatial extent of amorphous order, {\it i.e.} the length-scale over which particles (or molecules) in super-cooled liquids are arranged in an ordered, even though apparently chaotic, fashion. Numerical simulations have shown that this length $\xi_{PS}$, called point-to-set, grows upon super-cooling and plays an important role in the static and dynamical behaviours. In this work we unveil the existence of a second static length-scale, that together with $\xi_{PS}$ is central to the physics of super-cooled liquids and rules the relaxation within the RFOT picture.\\
The main physical ingredient of RFOT are the surface tension $\Upsilon$ and the configurational entropy density $s_c(T)$.  The former is a measure of the extra free-energy cost paid when two different amorphous phases are in contact through a common surface. The latter quantifies the multiplicity of amorphous phases in which the liquid can freeze. The mosaic state results from the competition between the configurational entropy gain due to local fluctuations between all possible amorphous states (all different type of "tiles" of the mosaic) and the surface energy loss due to the mismatch at the boundary between two amorphous states. A lot of analytical and numerical works have been devoted to characterise the spatial extent of the "tiles" of the mosaic, also called cooperative rearranging regions (CRR). Very few investigations have instead focused on their {\it interfaces}, for which no clear picture has arisen yet.    
Analytical studies based on Kac-models describe these interfaces as flat \cite{frasem11,zarfra10}, whereas numerical works instead suggest that these interfaces wander similarly to domain walls in disordered magnets \cite{bibcgv08,cacggv09a,cacggv09b}. Their width has been directly measured in recent numerical simulations and 
showed to grow mildly when temperature is lowered \cite{parisi09,berkob12,korobe12,sckobp02,grtcgv13,hobekr14}. The physical reason for this growth is still to be elucidated though. Basic questions remain unanswered: How much do these interfaces fluctuate? How do their fluctuations depend on temperature, in particular do interfaces become rougher or flatter 
approaching the glass transition? How does their width compare with the other characteristic length, the point-to-set length, which measures the spatial extent of CRRs? 
In order to answer all these questions,
fully characterise the real space structure of super-cooled liquids and the mosaic state advocated by RFOT theory it is crucial to develop a complete theory of fluctuating interfaces between amorphous states. This task is particularly timely since amorphous interfaces
have started to be directly probed in recent numerical simulations on pinned particle systems {{\cite{parisi09,berkob12,korobe12,sckobp02,grtcgv13,hobekr14}}}, and the first experimental results obtained by using optical traps in colloidal liquids have just come out \cite{gonags14}.\\
In this work we develop such a theory, obtain detailed predictions and provide explanations for previous numerical findings. We show that interfaces are rough because pinned by self-induced disorder and that they are characterised by wandering exponents identical to the ones of domain walls in the Random Field Ising Model (RFIM), thus strengthening the relationship between super-cooled liquids and the RFIM found in {{\cite{stewol08,frparr11,cambir13,frapar13,bicatt14}}}. {{These results allow us to establish that super-cooled liquids are characterized by two different static lengths, $\xi_{PS}$ and $\xi_\perp$, which  measure respectively the spatial extent and the 
fluctuations of the shape of CRRs. 
Their scaling with respect to $s_c$ is different: 
$\xi_{PS}\propto s_c(T)^{-1/(3-\theta)}$ with $\theta>1$, whereas $\xi_\perp\propto s_c(T)^{-1/2}$. Therefore, when $s_c$ decreases approaching the glass transition, boundary fluctuations grow but less strongly than the linear size. The resulting shape of CRRs is shown pictorially in Fig. 1.}} 

{{In order to study amorphous interfaces we adapt the theoretical protocol that has been used to characterize the size of CRRs \cite{boubir04}. We focus on equilibrated configurations, $\cal C'$, constrained to have a high overlap behind an infinite plane with a reference equilibrium configuration, $\cal C$. This can be operativly realised by taking an equilibrium configuration, pinning all particles behind a plane and resampling the configuration of the remaining free particles. The overlap field between $\cal C$ and $\cal C'$ is bound to be high close to the plane and to reach a low value, characteristic of bulk behavior, far from it. We define the {\it amorphous interface} for a given $\cal C'$ as the surface separating the high overlap region from the low overlap one.}}
{{Within RFOT theory, the physics behind the formation of amorphous interfaces is similar to wetting \cite{fisher84}: on the one hand it is favourable for the system to change amorphous state (or CRRs) beyond the plane because this allows a net gain in configurational entropy $s_c$. On the other hand, this leads to a free-energy loss due to the $\Upsilon$.
However, since the loss term scales as the surface, whereas the gain term scales as the volume, the former 
cannot counterbalance the latter and the drop in the
overlap field is always favourable. Understanding how this takes place and how the resulting amorphous interfaces fluctuate is one of the main aim of this work.}}
 {{An important remark on the procedure proposed above is that although $\xi_{PS}$ is finite, it holds $\xi_{PS}\gg\xi_\perp$ when $s_c$ is small, as we shall show. Thus, the curvature of CRRs is negligible on the scale over which interfaces fluctuate. On this scale the approximation of considering that the average profile of interfaces is flat and infinite, i.e. taking the overlap high behind an infinite plane, is justified. 
For simplicity, we shall first 
restrict our study to this case and 
take later into account the finite extension of CRRs. 
In the following we only focus on the temperature regime below the mode coupling transition temperature, $T_{MCT}$, where high and low overlap states are well-defined and it makes sense to use the concept of configurational entropy. 
We shall come back in the conclusion to the regime close to $T_{MCT}$.}}
\\
As noticed in several works, see 
\cite{frapar95,krzzdb11,cambir13} and refs therein, the overlap plays the role of an order parameter and the configurational entropy acts like a field favoring 
the low overlap state. In consequence, the situation is similar to the case of the ferromagnetic Ising model in {{a field, $H$,}} below the critical temperature. The counterparts of high and low overlaps are positive and negative magnetisations, whereas the configurational entropy plays the role of {{a negative field $H$}}. This analogy has been proved to be instructive in understanding the physics of glass-forming liquids \cite{krzzdb11,cambir13}. In consequence, we start our analysis by 
discussing the physical picture one obtains from it. Whithin this framework, pinning particles behind a wall  amounts to forcing all spins behind a wall to point up. In this way,
one induces an interface between the positively magnetized region close to the wall and the negative magnetized region favored by $H$ far from the wall. It is well known that in this case the effective Hamiltonian for the interface position $h({\mathbf x})$ ($h$ is the distance between the interface and the plane at position ${\mathbf x}$ in the $d-1$ planar dimensional space) reads {{\cite{dikrwa80}}}:
\begin{equation}
{\mathcal H}[h(\mathbf x)]=\int d{\mathbf x}\left[ \sigma \frac{\nabla h^2}{2}+Hh(\mathbf x)\right]
\label{Hising}
\end{equation}
where $\sigma$ is the surface-energy cost.
The statistics of the interface is obtained by integrating over all interface configurations weighted by their corresponding Boltzmann weight with the constraint $h{{(\mathbf x)}}\ge 0$. At zero temperature ${\mathcal H}$ is minimized by 
choosing $h{{(\mathbf x)}}=0$ for all $\mathbf x$, i.e. the interface is flat and stuck on the plane. 
For finite temperatures the interface fluctuate to gain entropy. In two dimensions the interface is a line and the functional integral can be mapped into a quantum mechanical problem that can be solved exactly \cite{fisher84}. One finds that is entropically favorable for the interface to wander over a length $\xi_{\perp}\propto H^{-1/3}$ perpendicular to the plane and a length 
$\xi_{\parallel}\propto H^{-2/3}$ parallel to the plane. 
In three dimensions the wandering is logarithmic only, and in four and higher dimensions the interface is flat \cite{fisher89}. 
In conclusion, the analogy with the ferromagnetic Ising model 
suggests that 3D amorphous interfaces are essentially almost flat and $\xi_\perp$ diverges logarithmically with $s_c$. A previous analysis based on Kac models also lead to a similar conclusion: interfaces are flat and characterized by $\xi_\perp \propto  -\ln s_c$ (for energetic reasons) \cite{frasem11,zarfra10}. 
However, a crucial physical ingredient 
has not been taken into account yet: self-induced quenched disorder. The specific reference configuration naturally introduces quenched randomness in the problem, which plays a very important role in the physics of super-cooled liquids as already shown in
\cite{stewol08,frparr11,cambir13,frapar13,bicatt14}. 
Note that in a super-cooled liquid there is no frozen-in disorder: it is the configuration from which the system has to escape in order to flow that plays the role of $\mathcal{C}$, i.e. of self-induced disorder.  
As it is known for random manifolds in random environments, disorder leads to a huge enhancement of the wandering of the interface, so large that thermal fluctuations 
become completely irrelevant. It is reasonable to expect that a similar phenomenon could also take place for amorphous interfaces. Indeed, by using replica field theory we show that the Hamiltonian governing the long-wavelength fluctuations of amorphous interfaces is given by (\ref{Hising}) plus a random potential term $\int d \mathbf x\, V_R\left(h(\mathbf x),\mathbf x\right)$, whose statistical properties are the same ones found for interfaces in the Random Field Ising Model. The scaling theory of the the RFIM then allow us 
to work out the behavior of the wandering length $\xi_\perp$, 
which we find to diverge as the square root of $1/s_c$ in three dimensions. In the following we derive the mapping to the RFIM. 
The resulting effective action for amorphous interfaces and the corresponding 
scaling theory for $\xi_\perp$ are presented in the next section. 
\begin{figure}
\centerline{\vspace{0.6cm}\includegraphics[width=.3\textwidth]{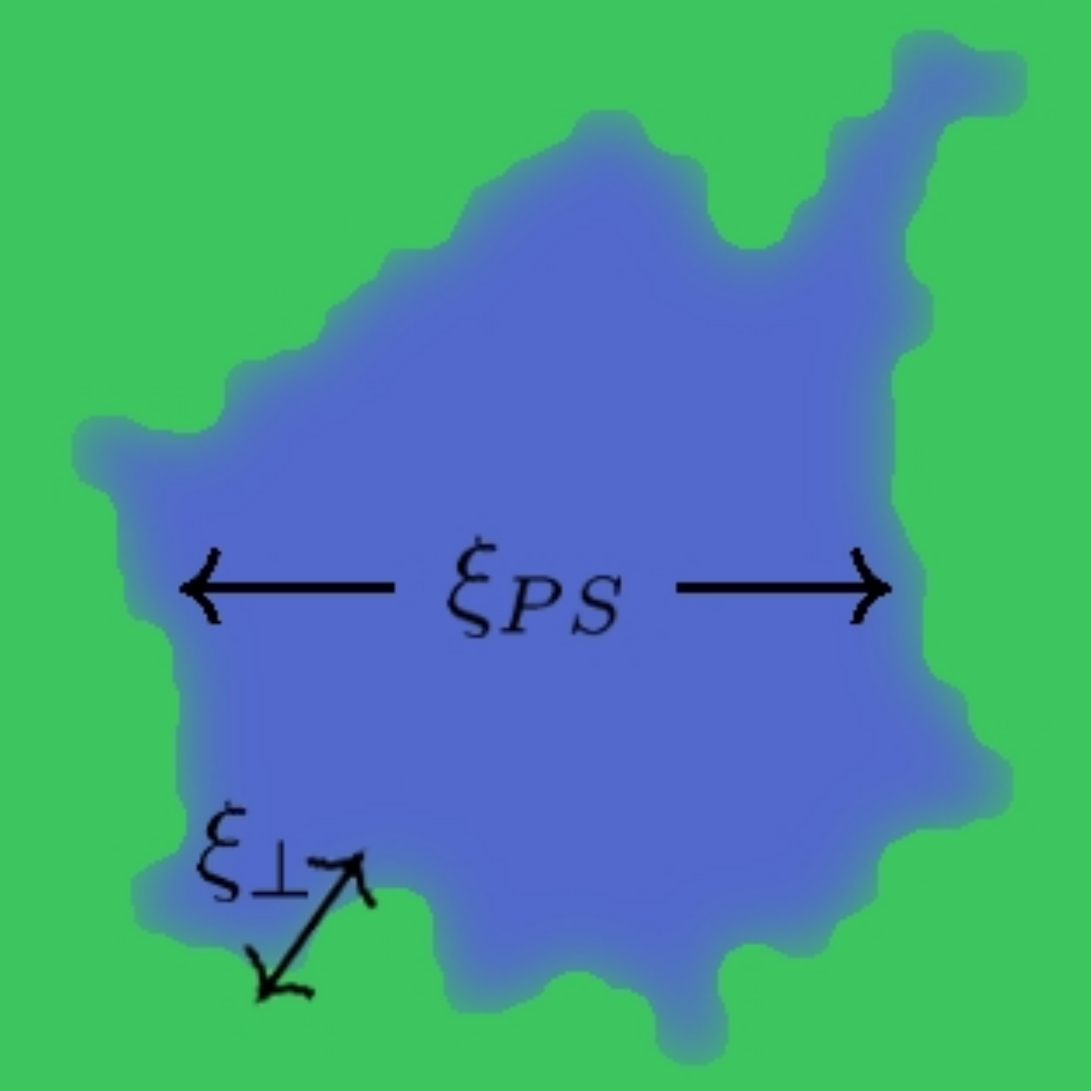}}
\caption{Cartoon of a cooperative rearranging region: the linear spatial extent is of the order $\xi_{PS}$ whereas the external shape is rough and fluctuating over the length-scale $\xi_\perp$.}\label{CRR}
\end{figure}
\section{Derivation and Mapping to the RFIM}
The starting point of our analysis is considering the
statistical field theory for the overlap field $p(z,{\mathbf x})$, which measures the similarity between two equilibrium configurations: the first is free whereas the second is constrained to coincide with the first behind a plane ($z$ is the distance from the plane and ${\mathbf x}$ are the coordinates along the plane). This boundary condition leads naturally to the existence of an interface, whose position 
along the $z$ axis, $h(\mathbf x)$, corresponds to the region in space where $p(z,{\mathbf x})$ jumps from the high value enforced by the constraint close to the plane to the low value favored by the configurational entropy. Our aim in the following is to obtain the effective field theory on $h(\mathbf x)$ starting from the one on $p(z,{\mathbf x})$. 
Following previous works that derived effective interface Hamiltonian \cite{dikrwa80} we assume that the most relevant configurations of $p(z,{\mathbf x})$ are the ones corresponding to a single interface positioned in $h(\mathbf x)$. This is natural since having more than one interface is unlikely. 
In consequence, henceforth we only focus on configurations $p^h(z,{\mathbf x})=q_{EA}$ for $0<z<h({\mathbf x})$ and zero for $z>h({\mathbf x})$, where $q_{EA}$ is the typical overlap of two configurations in the same amorphous state (it is 
associated with the local Debye-Waller factor characterizing molecular motion in the glass-forming liquid).  We neglect the smoothness of the decrease from one to $q_{EA}$ just after the wall and from $q_{EA}$ to zero at $z\simeq h({\mathbf x})$. Both simplifications are inessential to establish the effective field theory of $h({\mathbf x})$, as we shall discuss later. The effective Hamiltonian ${\mathcal H}_R[h(\mathbf x)]$ is obtained evaluating the action for the overlap field for $p=p^h(z,{\mathbf x})$: 
\[
{\mathcal H}_R[h(\mathbf x)]={\cal S}[p^h(z,\mathbf x)|{\cal C}]
\]
where we have made explicit the dependence on the reference equilibrium configuration ${\cal C}$ that introduces the quenched disorder. In order to show that ${\mathcal H}_R[h(\mathbf x)]$ coincides with  (\ref{Hising}) plus a 
random potential term, $\int d \mathbf x \,V_R\left(h(\mathbf x),\mathbf x\right)$, whose variance is the one characteristic of domain walls in the RFIM, we compute the average and the variance of ${\mathcal H}_R[h(\mathbf x)]$. We shall show that the former is equal to ${\mathcal H}[h(\mathbf x)]$ and the latter, $\overline{\left(({\mathcal H}_R[h_1(\mathbf x)]-{\mathcal H}[h_1(\mathbf x)])-({\mathcal H}_R[h_2(\mathbf x)]-{\mathcal H}[h_2(\mathbf x)])\right)^2}$ is {{proportional}} to the volume, ${\mathcal V_{h_1,h_2}}$, of the space embedded by the two interfaces described by $h_1(\mathbf x)$ and $h_2(\mathbf x)$. This is indeed the result expected for the RFIM case, where the correlator of the random potential {{is}} $\overline{V_R\left(h_1(\mathbf x),\mathbf x\right)V_R\left(h_2(\mathbf x),\mathbf x\right)} {{\propto}} |h_1(\mathbf x_1)-h_2(\mathbf x_2)|\delta(\mathbf x_1-\mathbf x_2)$ \cite{wiese03}.\\
The computation of the cumulants of ${\mathcal H}_R[h(\mathbf x)]$ 
is performed by introducing {{$n$}} different copies (or real replicas) of the system in presence of the same "disorder" $\cal C$ and averaging the replicated system over $\cal C$. Following Ref. {{\cite{bicatt14}}} we define the action of the replicated system by the identity
\[
\exp\left(-{\cal S}_r[\{p_a\}]\right)=\overline{\exp\left(-\sum_a {\cal S}[p_a(h,\mathbf x)|{\cal C}]\right)}^{\cal C}
\]
{{where $a\in[0,n]$}}.
The action ${\cal S}_r[\{p_a\}]$ generates all the cumulants of ${\cal S}[p(z,\mathbf x)|{\cal C}]$ through the equation:
\begin{equation}
{\cal S}_r[\{p_a\}]=\sum_a S_1[p_a]-\frac{1}{2}\sum_{a,b}S_2[p_a,p_b]+\frac{1}{3!}\sum_{a,b,c,}S_3[p_a,p_b,p_c]\dots
\end{equation}
where 
$$S_1[p]=\overline{{\cal S}[p|{\cal C}]}^{\cal C} \,\,,\,\,S_2[p_1,p_2]=\overline{{\cal S}[p_1|{\cal C}]{\cal S}[p_2|{\cal C}]}^{\cal C}-\overline{{\cal S}[p_1|{\cal C}]}^{\cal C}\overline{{\cal S}[p_2|{\cal C}]}^{\cal C}$$
and so on. \\
In order to extract $S_1[p]$ from ${\cal S}_r[\{p_a\}]$ one considers all replicas equal, i.e. $p_a=p\,\forall a$, and pick from ${\cal S}_r[\{p_a\}]$ the term linear in $n$ since $S_2, S_3,…$ are respectively of the order $n^2,n^3$ etc.. Similarly, in order to obtain $S_2[p_1,p_2]$ one subdivides all replicas in two groups, such that 
$p_a$ is equal to $p_1$ and $p_2$ for replicas respectively belonging to the first and the second group. 
In this case $S_2[p_1,p_2]$ is simply the part of the action proportional to $n_1n_2$ and can be therefore easily selected in the limit $n_1,n_2 \rightarrow 0$.  
The technical procedure to follow in order to compute $  {\cal S}_r[\{p_a\}]$ was derived in \cite{bicatt14}. In the following we just quote the final result: $  {\cal S}_r[\{p_a\}]$ is obtained as the free energy of the replicated field theory for $n+1$ copies $\alpha\in[0,n]$ of the system, in which one fixes the overlaps $q_{\alpha0}$ with the {{reference}} configuration {{labelled by}} $\beta=0$ to be equal to $p_a$
and integrates out all the others. As in Ref. \cite{bicatt14} the integration is performed by saddle-point (a more careful evaluation
of the functional integral is not expected to give rise to any qualitative change).   
We use as action of the replicated field theory the Landau's one proposed in Ref. {{\cite{dzscwo09}}}:
\begin{eqnarray}
{\cal S}[\{q_{\alpha\beta}\}]=\frac{E_0}{k_BT}\int_{z\mathbf x}\left\{\frac{c}{2}\sum_{\alpha\neq \beta}\left(\partial_{z\mathbf x}q_{\alpha\beta}(z,\mathbf x)\right)^2+\sum_{\alpha\neq \beta}v(q_{\alpha\beta})\right. \nonumber \\ 
\left. -\frac{u}{3}\sum_{\alpha\neq\beta\neq\gamma}q_{\alpha\beta}(z,\mathbf x)q_{\beta\gamma}(z,\mathbf x)q_{\gamma\alpha}(z,\mathbf x)\right\}
\end{eqnarray}
where it has been defined $v(q)=tq^2/2-(u+w)q^3/3+yq^4/4$, the main temperature dependence is in $t\sim k_B(T-T_0)/E_0$, $E_0$ is the liquid's energy scale, and $T_0$ is a constant.
From previous analyses we know that $v(q)+uq^3/3$ develops a secondary minimum below $T_{MCT}$ in correspondence of the Edwards-Anderson overlap value $q_{EA}$. The height of this minimum is the configurational entropy $s_c(T)$.\\
In order to compute the first cumulant we take $p_a=p^h(z,\mathbf x)$. Because of the 
resulting replica symmetry we solve the saddle point equations assuming $q_{ab}=q \ \ \forall a\neq b$ and $n\rightarrow 0$. We first consider the case of a flat interface, i.e. $h(x)=h$, $p^h(z,\mathbf x)=p(z)=q_{EA}\theta(h-z)$ where the latter is the Heaviside function.   
In this case $q$ only depends on $z$ and its saddle point equation reads: 
\begin{equation}\label{eqq}
c\partial^2_z q(z)-v'(q(z))-2uq(z)^2+uq_{EA}^2\theta(h-z)=0
\end{equation}
By numerically solving this equation\footnote{In the numerical integration the derivative of $q$ at the interface is chosen to optimize the action (actually maximize for reasons related to the $n\rightarrow0$ limit).} we found that an interface profile for $p^h$ induces
a similar interface profile $q^h(z)=I(z-h)$, where $I$ is constant until $z=h$ and then decreases rapidly to zero, see Fig. 1. This physically makes sense since replicas that are forced to have a high overlap with a reference configuration until a distance $h$ from the wall are expected to also have 
a high cross-overlap. 
By plugging these profiles into the replicated action and focusing on the term proportional to $n$ one naturally finds two contributions, one that is proportional to the volume between the wall and the interface and another one, independent on the interface position, that scales as the surface\footnote{In eq. (\ref{h1}) we only consider the leading term in $h$. The first correction to this behavior dependent on $h$ is of the form $\exp(-Kh)$ where $K$ is a constant. This term, in absence of quenched disorder, pushes the interface slightly away from the plane and leads to the result $\xi_\perp\propto -\ln s_c$ obtained in \cite{frasem11,zarfra10}. It can be neglected in the following since self-induced disorder leads to much stronger fluctuations and a more rapid increase of $\xi_\perp$ with $s_c$.}:   
\begin{eqnarray}\label{h1}
\overline{{\mathcal H}_R[h]}&\simeq&\int d{\mathbf{x}}\{ h[v(q_{EA})+uq_{EA}^3/3]+\Sigma\}\nonumber\\
&=&\int d{\mathbf{x}} s_c h+\Sigma L^{d-1}\,\,,
\end{eqnarray}
 where $\Sigma$ is the cost per unit surface of creating an interface and $L^{d-1}$ the wall surface. In order to compute $\Sigma$ correctly one should take into account an optimized and smooth form of the interface along the $z$ direction. We do not need to worry about this complication since the term 
linear in $h$, the one we are interested in, is independent of it.
Let us now also include in the analysis long-wavelength fluctuations of $h(x)$. Simple arguments show that in this case
$q(z,{\mathbf{x}})$ has to follow "adiabatically" the profile of 
$p^h(z,\mathbf x)=q_{EA}\theta(h({\mathbf{x}})-z)$, i.e. $q(z,{\mathbf{x}})=I(z-h({\mathbf{x}}))$ up to sub-leading corrections in gradients of $h({\mathbf{x}})$. By plugging this expression in the action one finally obtains that $\overline{{\mathcal H}_R[h]}$ at large length-scales and for $h\gg1$ is precisely equal to ${{\mathcal H}[h]}$ defined in eq. \ref{Hising}, with $\sigma=c$ and $H=s_c$, plus a constant term equal to $\Sigma$ times the wall surface. \\
Having obtained the first part of our technical results, 
we now turn to the study of the fluctuations of ${\mathcal H}_R[h]$. 
As discussed above, we have to consider two groups of replicas having 
an overlap profile with the reference configuration $p^{h_1}$ and $p^{h_2}$ respectively. As before we start by focusing on flat interfaces positioned at $h_1$
and $h_2$. Without loss of generality we will consider $h_1<h_2$.
By writing the saddle-point equation on $q_{ab}$ for $n_1,n_2 \rightarrow 0$ one finds that 
the overlap between replicas of the same group satisfies equation (\ref{eqq}) where the role of $p^h$ is played by $p^{h_1}$ and $p^{h_2}$ respectively. The overlap $q_{12}$ between replicas of different groups satisfies 
the equation:
\begin{eqnarray}
c\Delta q_{12}&=&v'(q_{12})-uq_{EA}^2\theta(h_1-z)\theta(h_2-z)\nonumber\\
&&+uq_{12}(q(z-h_1)+q(z-h_2)) \ .
\end{eqnarray}
Our numerical solutions \footnote{In the numerical integration the derivative of $q_{12}$ at the interface is chosen to optimize the action (actually minimize for reasons related to the $n\rightarrow0$ limit).} show that $q_{12}$ assumes a profile very similar to $p^{h_1}$, i.e. the interface profile closer to the wall: $q_{12}$ is equal to $q_{EA}$ for $z<h_1$ and has a sharp drop to zero just after, see Fig.2.
By plugging all the overlap profiles in the replicated action one finds that the 
second cumulant $S_2[p_1,p_2]$ reads
\begin{eqnarray}
S_2&=&-2\int_{z\mathbf x} \left\{ 2v(q_{12})-2up^{h_1}p^{h_2}q_{12}\right.
\nonumber \\
&&\left.+uq_{12}^2(q^{h_1}+q^{h_2})\right\} \ .
\end{eqnarray}
Using this result and in the limit $h_1\gg1$, $h_2\gg1$, $h_1-h_2\gg1$
we obtain that the variance of the fluctuations $\delta {\cal S}[p|{\cal C}]={\cal S}[p|{\cal C}]-\overline{{\cal S}[p|{\cal C}]}^{\cal C}$ reads:
\begin{eqnarray}\label{s2}
\overline{\left(\delta {\cal S}[p^{h_1}|{\cal C}]-\delta {\cal S}[p^{h_2}|{\cal C}]\right)^2}^{\cal C}&=&\\
&&\hspace{-2.8cm}=S_2[p^{h_1},p^{h_1}]+S_2[p^{h_2},p^{h_2}]-2S_2[p^{h_1},p^{h_2}] \nonumber \\
&&\hspace{-2.8cm}=4{\cal V}_{h_1,h_2} \left(\dfrac{u}{3}q_{EA}^3-s_c\right)\nonumber
\end{eqnarray}
{{where ${\cal V}_{h_1,h_2}$ is the volume embedded by the surfaces $h_1$ and $h_2$}}.
As discussed previously, if one considers long wave-length fluctuations of $h_1({\mathbf{x}}),h_2({\mathbf{x}})$, the overlaps $q^{h_1}, q^{h_2}, q_{12}$ follow adiabatically the solutions obtained for flat profiles. By plugging the corresponding solutions into the action one finds that $S_2$ acquires an extra contribution due to the gradient terms which can be neglected at leading order since scales as the area of the interfaces and not as the volume ${\cal V}_{h_1,h_2}$. In consequence, result (\ref{s2}) also holds for 
non-flat interfaces. 
\section{Effective Hamiltonian and Scaling Theory}
We now collect all previous results and write down 
the effective Hamiltonian governing the long-wavelength fluctuations of amorphous interfaces. It is a random functional of $h(\mathbf x)$ whose average and variance are the ones computed previously. We do not have computed higher cumulants but these are not expected 
to be relevant (in a renormalization group sense). Hence, for simplicity we shall take them equal to zero in the following. The final model for amorphous interfaces reads: 
\begin{equation}
{\mathcal H}_R[h(\mathbf x)]=\int d{\mathbf x}\left[ c \frac{\nabla h^2}{2}+s_ch(\mathbf x)\right]
+\int d \mathbf x \,V_R\left(h(\mathbf x),\mathbf x\right)+\Sigma L^{d-1}
\label{Hint}
\end{equation}
where $V_R(h(\mathbf x))$ is a random Gaussian potential with zero mean and $\Sigma$ is independent of the shape of the interface profile. The variance of $V_R(h(\mathbf x))$ reads
\begin{eqnarray}
\overline{V_R\left(h_1(\mathbf x_1),\mathbf x_1\right)V_R\left(h_2(\mathbf x_2),\mathbf x_2\right)}=\\
&\hspace{-2.8cm}=\left(\dfrac{u}{3}q_{EA}^3-s_c\right)|h_1(\mathbf x_1)-h_2(\mathbf x_2)|\delta(\mathbf x_1-\mathbf x_2)\nonumber
\end{eqnarray}
Note that the variance is positive, as it should, since in the regime we are interested in, i.e. $T$ close to $T_K$, the configurational entropy $s_c$ is small\footnote{It is possible to show that the disorder is short-ranged in $\mathbf x_1-\mathbf x_2$, thus for simplicity we have considered it $\delta$-correlated. Introducing explicitly finite range correlations would not alter our conclusions.}. \\
The model we ended up is identical to the one describing domain walls in the RFIM in presence of an external field. We can therefore use previous insights developed in this case, in particular the scaling theory of \cite{fisher89}, to work out the behavior of amorphous interfaces.   
The fluctuations of the interfaces are determined by the balance between two competing mechanisms.  
An interface closer to the wall leads to a gain of configurational entropy and hence to an {\it effective attractive interaction} $W_a(\ell)$. In fact the transition from high to the low overlap state at a distance $\ell$ from the wall leads to a free energy density gain (per unit surface) equal to $W_a(\ell)=W_a(0)+s_c \ell$. On the other hand, because of the random field disorder the interface wanders over increasingly large length-scales in order to find an optimized configuration 
that goes through favourable energetic regions, as it is known for the RFIM. Forcing the interface to wander no more than a distance $\ell$ from the wall induces a constraint 
and hence to a less optimized configuration, i.e. to a higher energy. As discussed in \cite{fisher89} this produces an {\it effective repulsive potential} between wall and interface (per unit surface) equal to $W_r(\ell)=W_r(0)+\frac{b}{\ell^\tau}$. The balance between these two mechanisms sets the value of the typical distance of the amorphous interface from the wall: $\xi_\perp\propto s_c^{-1/(\tau+1)}$. On length-scales smaller than $\xi_\perp$, the effective attraction due to the configurational entropy can be neglected and the interface fluctuations are similar to the ones of a free interface \cite{fisher89}. Thus, moving along the plane one encounters
over length-scales $\xi_\parallel\propto \xi_\perp^{1/\zeta}$ independent transverse fluctuations of the interface of the order $\xi_\perp$ ($\zeta$ is the roughness exponent of free RFIM interfaces).\\
For the RFIM the Imry-Ma argument, validated by Functional Renormalization Group analysis and numerical simulations, gives $\zeta=(5-d)/3$ and $\tau=2/\zeta-2=(2d-4)/(5-d)$ \cite{fisher89,wiese03}. This leads to a width of amorphous interfaces scaling as $$\xi_\perp\propto s_c^{-(5-d)/(d+1)}$$ In consequence, we finally find that 
amorphous interfaces wander  
in three dimensions over a length $\xi_\perp\propto s_c^{-1/2}$.
\begin{figure}
\centerline{\vspace{-.3cm}\hspace{-0.5cm}\includegraphics[width=.5\textwidth]{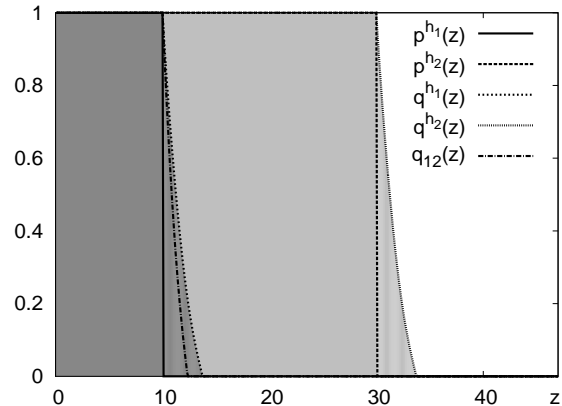}}
\caption{Overlap profiles $p^{h_1}(z)$, $p^{h_2}(z)$, $q^{h_1}(z)$, $q^{h_2}(z)$, $q_{12}(z)$ for $h_1=10, h_2=30$ and $s_c=0$ (the unit of length is $\sqrt{c}$). Note that as explained in the text $q^{h_1}(z)=I(z-h_1)$. For the Landau action considered in the text $q_{EA}=1$ when $s_c=0$. Similar results are obtained for a generic small value of $s_c$.}\label{profile} 
\end{figure}
\section{Physical consequences, predictions and comparison to numerical experiments}
Until now we have considered amorphous interfaces between regions of high overlap and low overlap separated by an infinite plane. This can be achieved by pinning particles 
behind a wall as discussed in the introduction (more on this later one). However, in a bulk glass-forming liquid the 
length which characterizes the spatial extent of amorphous order and the linear size of the CRRs is finite. It is called point-to-set and within RFOT scales as $\xi_{PS}\propto s_c^{-1/(d-\theta)}$ \cite{kithwo89}. The growth of $\xi_{PS}$ is due to a mechanism completely different from the one associated to $\xi_\perp$ and, accordingly, the growth law is different. In three dimensions, within a Kac-like (instanton) approach one finds $\theta=\left. d-1\right.=2$ \cite{franz05,dzscwo09}, scaling arguments by Kirkpatrick, Thirumalai and Wolynes suggest $\theta=\left.\frac d 2\right.= \frac 3 2$ \cite{kithwo89}, whereas some numerical results seem to indicate $\theta=2$ \cite{cacggv09a,cacggv09b}. Although a conclusive result on the value of $\theta$ is still missing, all indications point forward a value of $\theta$ such that 
 $1/(d-\theta)$, is larger than $1/2$ in three dimensions. 
In consequence, we do find as anticipated that CRRs, i.e. the regions over which the system is coherently in one amorphous state, are separated by interfaces that are rough
but fluctuate less than the typical size of the regions, see Fig. 1 for a pictorial representation. This remains true for higher dimensions. Interestingly, for $d=2$ the exponents of $\xi_{PS}$ and $\xi_{\perp}$ become equal ($d-1=d/2=(5-d)/(d+1)=1$) possibly indicating that $d=2$ is the lower critical dimension for the glass transition, as also suggested by other arguments {{\cite{cambir13,frapar13}}}.\\
The existence of two different static length-scales governing the physics of super-cooled liquids, $\xi_{PS}$ and  $\xi_\perp$, is a major fact to take into account
in understanding the outcomes of simulations and experiments in glass-forming liquids, in particular when probing static correlations. Our findings make clear that lengths extracted by different ways of pinning, in particular the wall geometry versus the spherical geometry, probe {\it different physical lengths}. 
This naturally provides an explanation for the difference found in numerical simulations between these two cases. In fact it was shown that the lengths probed using these two pinning geometries grow in a different way, more mildly in the former case as indeed expected from our results \cite{korobe12,grtcgv13}. Moreover, the decay of the overlap at the center of the cavity was found to be quite different from the decay of the overlap from the wall. In the first case, it becomes sharper at lower temperature (it can be fitted by an increasingly more compressed exponential) \cite{bibcgv08,homare12,grtcgv13}, whereas in the latter the form remains unchanged and exponential-like \cite{berkob12,grtcgv13}. These differences have a natural explanation within the physical picture arising from our theory in which two growing static lengths, $\xi_\perp$ and $\xi_{PS}$, intervene. Since the ratio $\xi_\perp/\xi_{PS}$ decreases by lowering the temperature, the CRRs
are better and better defined on the scale $\xi_{PS}$ and therefore the decay at the center of the cavity becomes indeed sharper and sharper at lower temperature. Instead, the decay of the overlap from the wall is governed by the length-scale $\xi_\perp$ only (the point-to-set length does not play any role). In this case it is natural to expect scaling with respect to $\xi_\perp$, as also shown for manifolds in random media, and hence 
a decay that does not change form, in particular does not become sharper by lowering the temperature.
Note that another case in which 
taking into account the existence of two different static length-scales
is crucial to explain numerical data has been discussed recently in \cite{macagr14}: in order to rationalize the finite size scaling of the specific heat for pinned systems in a cavity geometry 
one needs to consider both $\xi_\perp$ and $\xi_{PS}$. 
To test our scaling predictions it would be worth pushing further numerical simulations to obtain the dependence $\xi_\perp$ on $s_c$ for realistic model of super-cooled liquid. The RFIM character of amorphous interfaces that follows from our theory is already well supported by the numerical results of \cite{cacggv09a,cacggv09b} which found a roughening exponent $\zeta\simeq 0.62-0.75$ and energy fluctuations scaling as $\ell ^{2\zeta}$. These two results compare extremely well with our predictions, which also lead to the same scaling of energy fluctuations and  $\zeta=2/3\simeq 0.66$.
\section{Conclusion}
In this work we showed that amorphous interfaces are rough in three dimensions and we obtained the scaling with the configurational entropy of the length-scale over which they wander. 
Their statistical properties are identical to the ones of domain walls in random ferromagnets, a fact that strengthens even more the relationship between the physics of supercooled liquids and of the 
RFIM discussed in \cite{stewol08,frparr11,cambir13,frapar13,bicatt14}. 
One of our major results is that there are two different static length-scales governing the physics of super-cooled liquids: the point-to set length $\xi_{PS}$, related to the spatial extent of CRRs, and the wandering length $\xi_\perp$ 
related to the fluctuations of their external shape.
Our findings are in good agreement with previous numerical results, some of which were considered contradictory but find a natural explanation within our theory. 
We focused on the regime below $T_{MCT}$ where CRRs are well-formed and  configurational entropy and interfaces are meaningful concepts. Approaching $T_{MCT}$ we expect $\Upsilon$, and hence $c$, to decrease. This makes fluctuations 
more favorable. At a certain point,
when they become so large that the long-wavelength theory with a simple square gradient term is not suitable anymore, the 
description of the interface we used might break down. 
We suspect that close to $T_{MCT}$ this leads to different scaling forms and is associated to the 
fractal, or stringy, nature of CRRs found in {{\cite{stscwo06}}}. \\
Finally, in view of the recent studies of the glass transition in high spatial dimensions {{\cite{eavrei09,chikpz11,chikpz12,chcjpz13,chkupa14}}}
it is interesting to remark that our theory predicts a highly non-trivial dimensional dependence. In particular we find $\xi_\perp\propto s_c^{-(5-d)/(1+d)}$ and, hence, an upper critical dimension $d_u=5$. 
In higher dimensions amorphous interfaces are flat, $\xi_\perp$ does not increase and, hence, only one static growing length-scale accompanies the glass transition. This is a striking change in the nature of the glass transition that would be worth testing numerically.\\
In conclusion, the predictions obtained in this work 
provide a full characterization of the shape of cooperative rearranging regions in super-cooled liquids. They are instrumental in interpreting, understanding and devising new numerical simulations and experiments on static correlation in glass-forming liquids and clarify differences and relationships between the plethora of static lengths studied in recent years. An issue worth studying further, that we leave for future work, is whether in cases more accessible to experimental investigations, such as free surfaces instead of amorphous walls, the length $\xi_\perp$ can be probed \cite{gruebel10}. 


\begin{acknowledgments}
We thank T. Giamarchi, G. Parisi, G. Tarjus and M. Tarzia for very useful discussions. 
We acknowledge support from the ERC grants NPRGGLASS (GB) and CRIPHERASY (CC) (no. 247328).  
\end{acknowledgments}
\bibliographystyle{unsrt}
\bibliography{bib}

\end{article}

\end{document}